\documentstyle[psfig]{caosp}

\def\DA{$\Delta$a }

\begin{document}

\title{CP2 stars in clusters: deep $\Delta a$-photometry}
\author{H.M.~Maitzen, M.~Rode, E.~Paunzen}
\institute{Institut f\"ur Astronomie der Universit\"at Wien,
         T\"urkenschanzstr. 17, A-1180 Wien (surname@astro.ast.univie.ac.at)}
\maketitle

\begin{abstract}
The search for chemically peculiar (CP) stars in open clusters using
photoelectric photometry sampling the presence of the characteristic
flux depression feature at 5200\AA\ via the $\Delta a$-system (Maitzen 1976)
has so far delivered data for objects usually no more distant than 1000 pc
from the Sun. If one intends to study the presence of CP stars
at larger distances from the Sun, classical photometry has to be
replaced by CCD photometry.

For the first time, our investigation presents the results of CCD-photometry
in the $\Delta a$-system for a rich open cluster which is at a distance clearly
beyond hitherto studied objects, Melotte 105 (2\,kpc, log age = 8.5).

Comparison with published $uvby$-photometry yields the
calibration of the colour index $g_{1}-y$ of our system, which is necessary for
deriving the peculiarity index $\Delta a$. For this we achieve an average
accuracy of 0.007 mag.

Six objects with only marginally peculiar $\Delta a$-values were found, but 
spectroscopic and additional photometric evidence is needed to substantiate
their peculiarity.
\end{abstract}

\section{Introduction}

Chemically peculiar (CP) stars of the upper main sequence have been targets
for astrophysical studies for exactly 100 years. Starting with the pioneering
work of Maury (1897), most of this interval was devoted
to the detection of peculiar features in their spectra and photometric 
behaviour, leading to the post World War II discovery of magnetic fields
 and the Oblique Rotator concept of
slowly rotating stars with non-coincidence of magnetic and rotational axes.
Since spectroscopic work required high resolution to accomplish magnetic field
measurements, the determination of small Doppler
shifts and the identification of the highly crowded narrow lines, 
investigations 
of CP-stars tended rather to the study of exotic individual behaviour than to
the extraction of features common to this special group of main sequence stars.

Only 30 years ago a statistical study using the Str\"omgren system paved
the way to a period of work aimed at shaping our concept towards questions
on the origin of CP-stars, their evolution on the main sequence and their 
dependence on other parameters like metallicity, location in the galactic plane
etc.

The prerequisite for investigating larger samples of CP-stars 
(including the generally fainter open
cluster members) is unambiguous detection. Looking into catalogues of CP-stars,
especially of the magnetic ones, it immediately becomes obvious that there are
many discrepancies at classification dispersion. Even the short list of
peculiar stars identified by Maury (1897) contains one object which 
is classified as erroneous entry in the catalogue of Renson (1991).

The reasons for discrepant peculiarity assessments are to be found in the
differences of personal pattern recognition among different classifiers,
differences of (mainly photographic) observing material (density of
spectrograms, widening of spectra, dispersion, focussing),
 seeing conditions (for
objective prism spectra), but also intrinsic variability of
peculiar spectral features (e.g. silicon lines).

Photometry has shown a way out
of this dilemma, especially through the discovery of characteristic broad band
absorption features, the most suitable of them located around 520 nm. Two
decades ago Maitzen (1976)
introduced a 3 filter photometric system which samples the depth of this green
flux depression by comparing the flux in the center (522 nm, $g_{2}$), with the
adjacent regions (500, $g_{1}$, and 550 nm, $y$) using a band-width of 13 nm.
The respective index was introduced as:
\begin{center}
    $a=g_{2}-(g_{1}+y)/2$
\end{center}
Since this quantity is slightly dependent on temperature (increasing towards
lower temperatures), the intrinsic peculiarity index had to be defined
as 
\begin{center}
    $\Delta a= a-a_o[b-y]$
\end{center}
i.e. the difference between the individual a-value and the a-value of
non-peculiar stars of the same colour (the locus of $a_o$-values has
been called {\it normality line}).

Maitzen (1976) showed that virtually all peculiar stars with magnetic fields,
the CP2-group according to the definition of Preston (1974), have positive
$\Delta a$ values larger than 0.012 mag up to 0.075 mag.

This result was corroborated by Maitzen \& Vogt (1983)
for a significantly larger sample.
In a series of papers (first:  Maitzen \& Hensberge 1981; last: Maitzen 1993)
38 open clusters have been surveyed for the presence of magnetic
peculiar stars using $\Delta$a-photometry with the conventional photomultiplier
technique. The advantage of this approach is both a relatively high accuracy
of the detection index $\Delta$a (external scatter for normal stars 3-5 mmags)
and its immediate availability, hence detection of CP2-stars right at the
telescope (provided that the data acquisition system is able to display
linear combinations of on-line magnitudes).

On the other hand the magnitude limit is about 12 for work with a 1m telescope.
This limit corresponds to a vicinity of only about 1 kpc around the Sun. Even
in this area we can reach in some cases only the hotter domain of CP2-stars.

If one considers to study the behaviour and appearance of peculiar stars
over a significant range of galactocentric distances - speculating about
the possible influence of different degrees of metallicity and/or
galactic magnetic field strengths - then it is imperative to increase the
actual magnitude limit by at least 2-3 magnitudes. 
Discarding the use of 4m telescopes as highly unrealistic for this purpose we
are left with employing the CCD-technique. Maitzen et al. (1997) have
prepared the ground by showing that for field stars in the range 8$<$V$<$10,
$\Delta a$ values obtained with CCD-photometry at a 60cm telescope exhibit
essentially the same error level as conventional photoelectric photometry.

Aside from the basic advantage of CCD-photometry - i.e. simultaneous
recording of many individual objects not only saving time but also
increasing precision - the second benefit becomes more and more obvious
when we observe clusters at larger distances with decreasing angular
separations of their members, and this is the determination of
stellar magnitudes by fitting of point spread functions.

The primary goal of the present study is to evaluate the accuracy
performance of CCD-photometry, i.e. to which limiting magnitude
the peculiarity index $a$ viz. $\Delta a$ can be determined in an open
cluster with
a precision comparable to classical photoelectric photometry
within reasonable amounts of observing time.

Here we present the results for our first open cluster, Melotte 105,
 measured in the $\Delta a$ system
with the Bochum 61cm telescope during a campaign in April/May 1995 at
 ESO La Silla,
during which  3 dozens of other open clusters were observed.

Mel 105 is at a distance of 2000 pc in the 4th galactic quadrant and
has a reddening of $E(B-V)=0.38$. Therefore, it is at a larger distance
than all other open clusters surveyed so far by photoelectric photometry
in $\Delta a$ and a good starter for studying remote clusters in this
system.

\section{Observations and Reduction}

Observations were performed with the 61 cm Bochum telescope at ESO-La Silla on 
4 photometric nights (April 8, 9, 13, 16, 1995).
 The telescope was equipped with a 
liquid nitrogen cooled Thompson 7882 CCD (384x576 pixels) resulting in a field
 of view of about 3' by 4'.
Three Schott interference filters (diameter of 50mm) were used.
Two frames per filter were obtained on each of the afore-mentioned nights;
the integration time was always 600 seconds per frame. 

The basic CCD-reduction steps (bias-subtracting and flat-fielding) were
performed using standard IRAF-routines. Contrary to our introductory paper
on CCD-$\Delta a$ photometry (Maitzen et al. 1997),
we had to resign about aperture photometry, and used
the package DAOPHOT fitting point spread function profiles to the
stellar images. 

Our sample contains 71 stars brighter than $V=15$ which exhibit an
average scatter of 0.0040 mags around their mean $a$-values. The remaining
43 stars up to the magnitude limit 16 in $V$ scatter by 0.0073 mags
around their $a$-averages.

In order to establish the line of normality for deriving the
deviations \DA, we have chosen
{\it reference stars} according to the following
criteria: \\
They had to exhibit an average behaviour both in the $m_{1}$ vs. $b-y$ and
$c_{1}$ vs. $b-y$ diagrams based on the CCD-photometric study of Balona \&
Laney (1995). This way 16 objects
have been obtained spanning an interval in $b-y$ from 0.28 to 0.50
corresponding to B9 - A8 stars after subtraction of the cluster reddening.

As abscissae in the $a$ vs. colour diagrams we were able to choose among
3 types of temperature indices: \\
photographic $B-V$ values from Sher (1965), $b-y$ just mentioned and
$g_{1}-y$ of our own photometric system. 
From the correlation of our index with either
$B-V$ and $b-y$, we immediately conclude that we should resign
about using the photographic $B-V$ data for deriving $\Delta a$-values.
On the other hand it is very comforting to notice the very good
correlation of CCD-based $b-y$ and $g_{1}-y$.

Therefore we decided to derive $\Delta a$-values from normality lines
based on both $b-y$ and $g_{1}-y$:
$$ a_{0} = 0.603  + 0.139\,(b-y) $$
and
$$ a_{0} = 0.753 + 0.197\,(g_{1}-y) $$
The $\Delta a$-values scatter around their lines by 0.0028
and 0.0034 mags, respectively.
Assuming that the $\Delta a$ values of the reference stars 
follow a normal distribution, we have calculated confidence intervals
(Rees 1987) of 99 percent, respectively. 

The average scatter of the $\Delta a$ values derived from the $b-y$
normality line is 0.0073 if we regard only those 45 stars which are
brighter than $V=15$. With the $g_{1}-y$ normality line we get a scatter
of 0.0079 for the 68 stars of the same brightness interval.
Had we taken all objects in both cases we would have arrived at mean
scatter values of 0.0088 and 0.0104 for $b-y$ and $g_{1}-y$ , resp.
The reason for the larger difference in the second case is the
lower number of stars fainter than $V=15$ measured by Balona \& Laney (1995).
Fainter stars do not only increase the scatter in $\Delta a$ because of
photon statistics, but also because of the higher reddening usually
related to background stars which moves them to the right and away from the
normality line.

The opposite case - foreground stars with lesser reddening - will also
increase the $\Delta a$ scatter, shifting the points systematically
to the left, even leading to the possibility of mimicking a peculiar star.

We conclude that the assignment of peculiarity is less straightforward in the
case of clusters more distant than those
so far observed with conventional photoelectric $\Delta a$ photometry.

Six objects could be considered as
mild photometric CP2-candidates which ought to be verified by
spectroscopy. But none of
them could be called an outstanding photometric peculiar object, since
the positive deviations from the normality line hardly reach 0.020 mag.

\section{Results}

Melotte 105, a rather rich open cluster 
with more than 100 stars measured in our programme does not possess
any pronounced photometrically peculiar star of type CP2, although it has
many stars around A0 where the relative frequency of those stars
reaches its maximum. The probability to identify a mild peculiar
star from 6 candidates by spectroscopy does not seem very high.
Our sample shares 68 stars in common with that observed in CCD-photometry by
Balona \& Laney and the high degree of correlation between their
$b-y$ values and our $g_{1}-y$ values justifies the use of the latter ones
for the purpose of deriving $\Delta a$-values.
The larger negative deviations for the fainter stars of our sample
(which are significantly less present in the Str\"omgren study) can,
therefore, be easily ascribed to the effect of larger interstellar
reddening of background stars.

As a result of this investigation, we state that for the observing
time invested (240 minutes of net integration time) for this cluster,
we achieve a threshold value for peculiarity slightly above 0.020 mags
for stars brighter than $V=15$. Therefore we conclude that about
twice the amount of integration time would be sufficient to lower
the threshold to values obtained with classical photoelectric
photometry (with $V=11$ as the limit for the same telescope size!). 

The big step forward in saving telescope time becomes visible if one
considers that for a cluster with the same amount of stars, but at
the $V=11$ level we would have needed 50 hours of photoelectric 
observing at the same telescope!

\acknowledgements
HMM acknowledges financial support from the Hochschuljubil\"aumsstiftung
der Stadt Wien (project {\it Wiener Zweikanalphotometer}).
EP acknowledges support of this research  within the working group
{\it Asteroseismology-AMS} with funding from the Fonds zur F\"orderung
der wissenschaftlichen Forschung (project {\em S7303-AST}).
Use was made of the Simbad database, operated at CDS, Strasbourg, France.

\end{document}